\documentclass[runningheads]{llncs}

\usepackage{multirow}
\usepackage{bbding}
\usepackage{amsmath}
\usepackage{comment}
\usepackage{hyperref}
\usepackage{url}

\usepackage[T1]{fontenc}
\usepackage{graphicx}
\begin{document}
%

\title{Unsupervised Domain Adaptation for Pediatric Brain Tumor Segmentation}

%


\author{Jingru Fu\inst{1}\orcidID{0000-0003-4175-395X} \and
Simone Bendazzoli\inst{1, 2}\orcidID{0000-0001-6673-1314} \and
Örjan Smedby\inst{1}\orcidID{0000-0002-7750-1917} \and
Rodrigo Moreno\inst{1}\orcidID{0000-0001-5765-2964}}
\authorrunning{F. Author et al.}

\institute{KTH Royal Institute of Technology, Stockholm, Sweden\\
\email{\{jingruf,simben,orsme,rodmore\}@kth.se}\\ \and
Karolinska Institutet, Stockholm, Sweden\\
\email{simone.bendazzoli@ki.se}}
\maketitle              
\begin{abstract}
Significant advances have been made toward building accurate automatic segmentation models for adult gliomas. However, the performance of these models often degrades when applied to pediatric glioma due to their imaging and clinical differences (\textit{domain shift}). Obtaining sufficient annotated data for pediatric glioma is typically difficult because of its rare nature. Also, manual annotations are scarce and expensive. In this work, we propose Domain-Adapted nnU-Net (DA-nnUNet) to perform unsupervised domain adaptation from adult glioma (\textit{source domain}) to pediatric glioma (\textit{target domain}). Specifically, we add a domain classifier connected with a gradient reversal layer (GRL) to a backbone nnU-Net. Once the classifier reaches a very high accuracy, the GRL is activated with the goal of transferring domain-invariant features from the classifier to the segmentation model while preserving segmentation accuracy on the source domain. The accuracy of the classifier slowly degrades to chance levels. No annotations are used in the target domain. The method is compared to 8 different supervised models using BraTS-Adult glioma  (N=1251) and BraTS-PED glioma data (N=99).
The proposed method shows notable performance enhancements in the tumor core (TC) region compared to the model that only uses adult data: $\sim$32\% better Dice scores and $\sim$20 better 95th percentile Hausdorff distances. 
Moreover, our unsupervised approach shows no statistically significant difference compared to the practical upper bound model using manual annotations from both datasets in TC region.
The code is shared at \url{https://github.com/Fjr9516/DA_nnUNet}.

\keywords{Unsupervised domain adaptation  \and Pediatric tumor segmentation \and Gradient reversal layer.}
\end{abstract}
\section{Introduction}
Pediatric brain tumors represent the most prevalent solid tumors and the leading cause of cancer-related mortality in children \cite{fathi2023automated}. Accurate segmentation of these tumors from medical images is critical for surgical and treatment planning \cite{menze2014multimodal,fathi2023automated}, facilitating the identification of tumor location, extent, and type. 

Recent advances in automated brain tumor segmentation, driven by deep learning techniques and well-curated datasets, have shown promise in adult brain tumors \cite{isensee2021nnu,luu2021extending,zeineldin2022multimodal}. In turn, an important issue for pediatric brain tumors is the scarcity of training data. For instance, while the incidence of adult glioblastomas is 3 in 100,000 people, pediatric diffuse midline gliomas are approximately three times rarer \cite{kazerooni2023brain}. An additional issue is that obtaining accurate segmentation models 
involves expert annotation of data, which entails laborious processes. 

An approach to deal with this data scarcity issue is to train deep-learning models with adult data and apply them to pediatric cases. This approach relies on the assumption that brain tumors are similar in both adults and children. Unfortunately, this approach often fails to generalize effectively to pediatric cases. For example, \cite{zeineldin2022multimodal} reported a sharp decrease in the Dice similarity coefficient score (DSC) for the tumor core (TC) region, dropping from 0.8788 to 0.2639, when applying a model developed for adult gliomas to pediatric cases.
Indeed, despite certain similarities between pediatric and adult tumors, there are also important differences in their appearance (see an example in Fig.~\ref{fig:domaindiff} (right)).
For instance, enhancing tumor regions on post-gadolinium T1-weighted MRI and radiologically-defined necrotic tissue regions are common in adult gliomas but are less common in pediatric cases \cite{kazerooni2023brain}. 

International challenges, such as the brain tumor segmentation challenge (BraTS) \cite{menze2014multimodal}, have been crucial for the development of successful approaches for adult tumor segmentation. 
BraTS provides participants with datasets of fully annotated, multi-institutional, multiparametric MR images (mpMRIs) of patients. Notably, winners of recent competitions have predominantly employed U-Net architectures, which demonstrate their superior performance, particularly for supervised learning with abundant training data. Among these, nnU-Net \cite{isensee2021nnu}, a self-configuring framework that automatically adapts to specific datasets, has gained popularity for its robust performance \cite{vossough2024training,kharaji2024brain}. Unsupervised and semisupervised approaches are promising for addressing problems with fewer training cases and/or with low to no annotation data, which is often the case of pediatric tumor segmentation.

\textbf{Contributions}: In this paper, we propose an unsupervised domain adaptation (UDA) method for pediatric glioma segmentation. In particular, we add a domain-adversarial task to a backbone nnU-Net. The goal of this task is to encourage the nnU-Net to learn domain-invariant features for segmentation. The inclusion of a domain classifier stems from principles of domain adaptation, which suggest that effective domain transfer requires predictions based on features that do not distinguish between the source and target domains \cite{ben2006analysis}. This concept was initially introduced in neural architectures in \cite{ganin2016domain} for natural image classification tasks, employing a gradient reversal layer (GRL) to learn domain-invariant intermediate features for the final task. Similar efforts have been made to extend this concept to natural image segmentation \cite{bolte2019unsupervised} and 3D medical binary segmentation problems \cite{kamnitsas2017unsupervised}. In this paper, we adapt this approach to the 3D region-based brain tumor segmentation problem, which, to the best of our knowledge, has not been previously explored.



\section{Method}
In domain adaptation (DA), the objective is to learn a model from a source domain $\mathcal{D}_s$ and apply it to a related but different target domain $\mathcal{D}_t$ for the same task ($\mathcal{T}_s = \mathcal{T}_t$). 
In this study, we focus on UDA and use supervised domain adaptation (SDA) to establish upper-bound and baseline models.


\begin{figure}[!t]
\includegraphics[width=\textwidth]{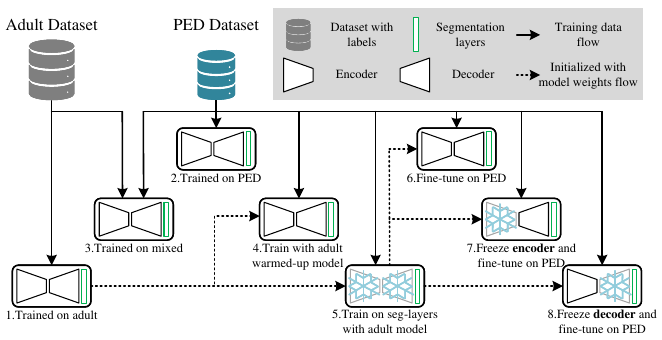}
\caption{Different strategies to obtain upper-bound and baseline models when assuming fully-annotated pediatric (PED) data are provided.} \label{fig:SDA}
\end{figure}
\subsection{Supervised Baseline Models}\label{sec:SDA}

Despite their differences, adult and pediatric tumors share relevant characteristics. Thus, using abundant annotated adult data during training should be beneficial for pediatric tumor segmentation \cite{boyd2023expert}.
In case manual annotations of pediatric images are available, a standard procedure is to apply transfer learning from adult to pediatric gliomas by combining the two datasets during training \cite{goodfellow2016deep,amod2023bridging}. This combination can be performed in different ways. 

Fig. \ref{fig:SDA} shows the 8 different supervised models. Models 1-3 in Fig. \ref{fig:SDA} train 3D nn-UNets from scratch using either or both datasets. In turn, models 5-8 use standard transfer learning strategies. First, model 5 is trained only on task-related layers, based on the pre-trained nn-UNet for adult tumor segmentation as the feature extractor. Then different parts of the network are fine-tuned with pediatric data, with lower learning rates and fewer epochs in models 6-8. Model 4 is somewhat in the middle of the two approaches, where the whole network is retrained using pediatric data after pretraining with adult data. 



\subsection{Unsupervised Domain Adaptation for Pediatric Brain Tumor Segmentation}
\begin{figure}[!t]
\includegraphics[width=\textwidth]{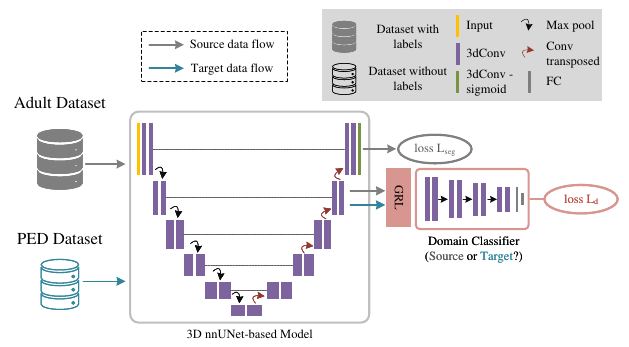}
\caption{The proposed DA-nnUNet comprises a 3D nnUNet-based backbone and a domain classifier. UDA is facilitated by integrating a gradient reversal layer (GRL) before the domain classifier. 
} \label{fig:UDA}
\end{figure}
As illustrated in Fig. \ref{fig:UDA}, the proposed architecture consists of a 3D nnUNet-based backbone and a domain classifier. During forward propagation, the GRL \cite{ganin2016domain} acts as an identity function, while during backpropagation, it multiplies the gradient by a negative value $\alpha$. This adversarial mechanism encourages the features passed to the domain classifier to become domain-invariant. At the final stage of the UNet decoder, which is responsible for the segmentation task, the segmentation loss $L_{seg}$ is minimized only on adult samples, while the domain classifier minimizes a domain classification loss $L_{d}$ across both domains.

 
The success of achieving domain adaptation with our model hinges on carefully balancing model learning between segmentation and domain classification tasks. Overfitting to the segmentation task may hinder domain adaptation, causing the model to excel only in source domain segmentation. Conversely, overfitting to the classification task may compel the model to prioritize learning domain-invariant features to deceive the classifier, potentially sacrificing segmentation performance. To address these concerns, we introduced several strategies to better balance the two tasks.

\textbf{Inclusion of domain classifier weight in loss}: The total loss comprises two primary components: a segmentation loss $L_{seg}$ and a domain classifier loss $L_d$. We employed a combination of pseudo-Dice loss and binary cross-entropy loss as $L_{seg}$, akin to the default loss function in nnU-Net. For $L_d$, we utilized cross-entropy loss for the one-hot with two domain labels. The total loss is represented as follows:

\begin{equation}
    L_{total} = \sum_{x\in \mathcal{D}_s} L_{seg}(\hat{y}, y) + \lambda \sum_{x\in (\mathcal{D}_s \cup \mathcal{D}_t)} L_{d}(\hat{d}, d),
\end{equation}
where $\lambda$ denotes the domain classifier weight, controlling the relative importance of the segmentation and classification tasks; $\hat{d}$ and $d$ are the predicted and actual domain labels; and $\hat{y}$ and $y$ represent the predicted and ground truth segmentation masks for the adult dataset $\mathcal{D}_s$, respectively. We assume no annotations are available for the pediatric dataset $\mathcal{D}_t$, so they are not considered in $L_{seg}$. 

\textbf{Dynamic scheduler for $\alpha$ in GRL}: A crucial component is the adjustment of $\alpha$ in the GRL, which dictates the strength of feature adaptation in response to the domain classifier. Following \cite{kamnitsas2017unsupervised}, we implemented a truncated ramp function in which the value of $\alpha$ grows linearly from 0 at epoch $e_{min}$ to $\alpha_{max}$ at epoch $e_{max}$. $e_{min}$ is selected such that the accuracy of the classifier is very high. Ideally, such an accuracy should decay to a chance level at $e_{max}$. Finally, we maintain $\alpha$ at its maximum value $\alpha_{max}$ to focus on refining further the segmentation.



\section{Experimental Setting}
\subsection{Datasets}

\begin{table}[t]
\caption{Datasets used in the experiments.}\label{tab:dataset}
\begin{tabular}{ccccc}
\hline
\textbf{Dataset} & \textbf{Data Subset} & \textbf{\# Subjects} & \textbf{\begin{tabular}[c]{@{}c@{}}Labels Used\\ (Supervised)\end{tabular}} & \textbf{\begin{tabular}[c]{@{}c@{}}Labels Used\\ (Unsupervised)\end{tabular}} \\ \hline
BraTS Adult Glioma & $\mathcal{D}_{\text{Adult}}^{\text{source}}$ & 1251 & \Checkmark & \Checkmark \\
BraTS-PEDs         & $\mathcal{D}_{\text{PED}}^{\text{target}}$ & 99   & \Checkmark &  \\
\hline
\end{tabular}
\end{table}

\begin{figure}[t]
     \centering
         \centering
         \includegraphics[width=0.48\textwidth]{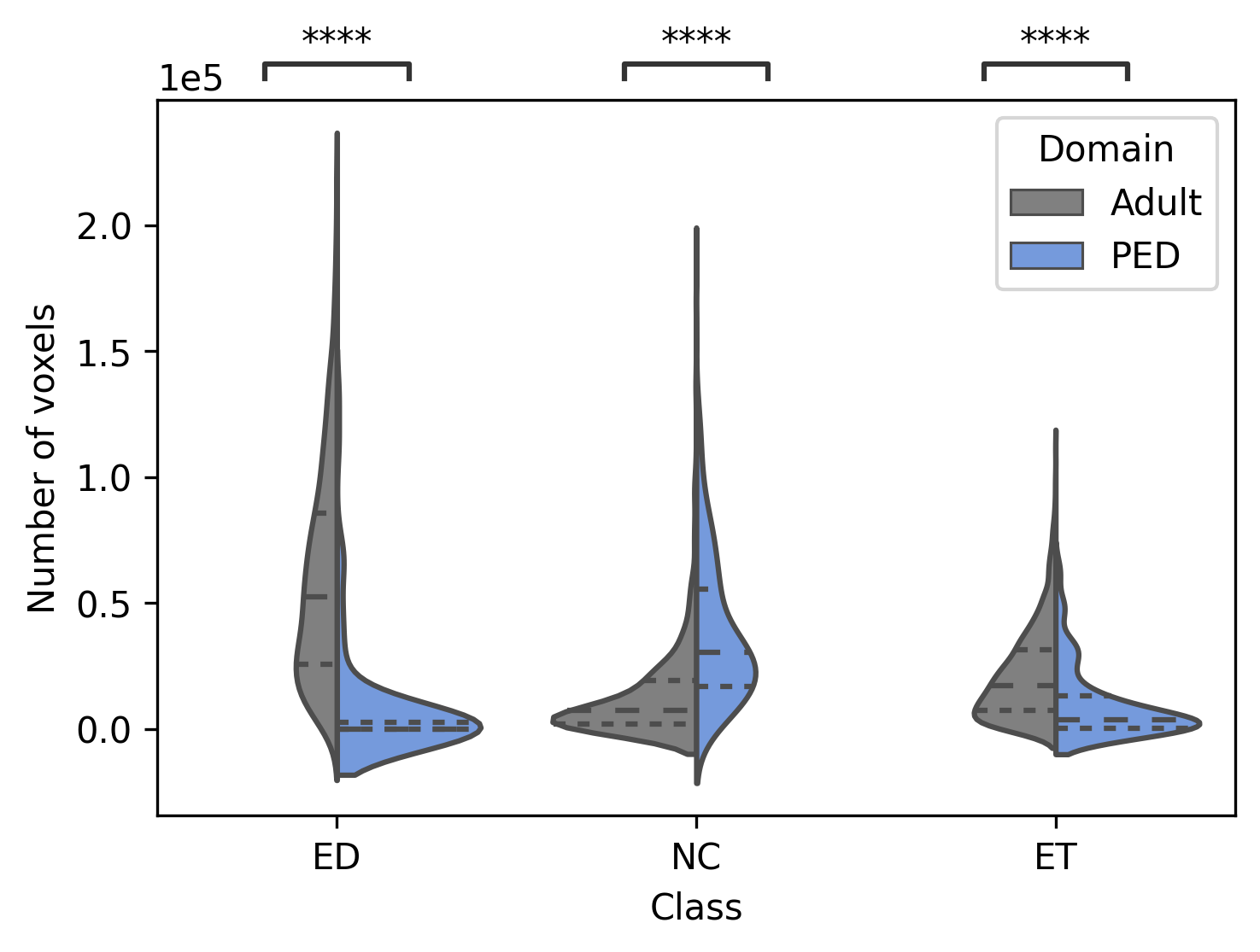}
     \hfill
         \centering
         \includegraphics[width=0.48\textwidth]{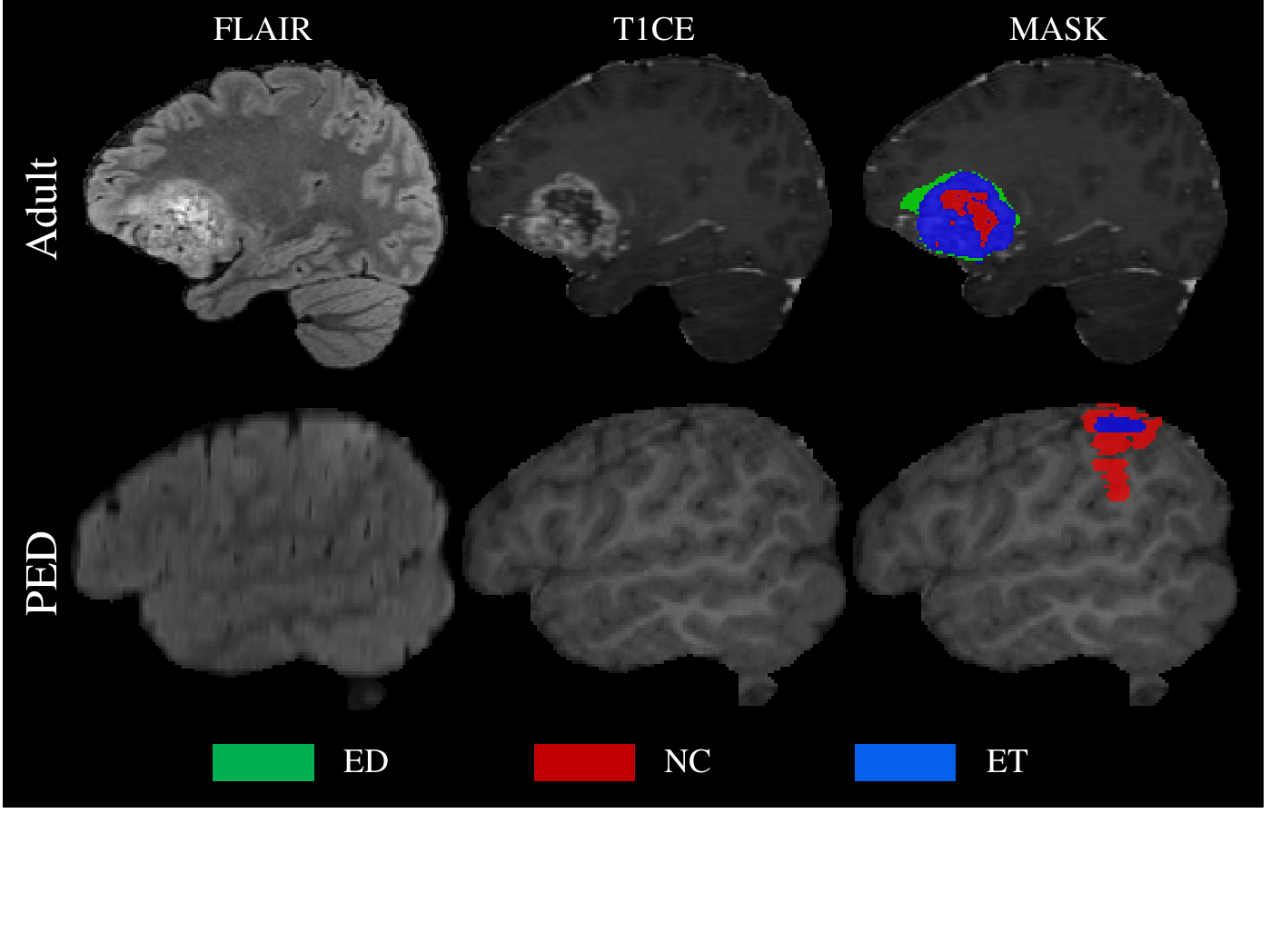}
\caption{Distribution of tumor sizes (left) and examples of the appearance of adult and pediatric tumors on FLAIR and T1w after contrast (T1CE) (right).
} \label{fig:domaindiff}
\end{figure}

We utilized two databases provided by the BraTS challenge organizers: the BraTS 2021 adult glioma dataset \cite{menze2014multimodal}, and the BraTS-PEDs 2023 dataset \cite{kazerooni2023brain}. All data underwent identical preprocessing steps, including conversion to NifTI format, skull stripping, defacing, co-registration to the SRI24 anatomical template, and resampling to isotropic resolution. Subsequently, the data were cropped to a size of [240, 240, 155]. Each dataset included four mpMRIs: pre- and post-gadolinium T1-weighted (T1w), T2-weighted (T2w), and T2 fluid attenuated inversion recovery (FLAIR) images. Additionally, expert annotated segmentation masks were provided, delineating three classes: enhancing tumor (ET), non-enhancing component (NC), and edema (ED).

We employed the training datasets to validate our method for the segmentation of three subregions: whole tumor (WT: ET+NC+ED), tumor core (TC: ET+NC), and enhancing tumor (ET). Details regarding each domain are summarized in Table \ref{tab:dataset}. All models using PED data were trained using 5-fold cross validation (CV), while \textit{adult} model was trained using all adult data. 
As shown in Fig. \ref{fig:domaindiff}, there are significant disparities (i.e., \textit{domain shift}) between adult and pediatric gliomas for tumor subregion sizes and appearance. Note that in the BraTS-PEDs set, 11 out of 99 cases lack the ET region (compared to 33 out of 1251 cases in the BraTS adult glioma set), while 42 out of 99 cases lack the ED region (compared to only 1 out of 1251 cases in the BraTS adult glioma set). 

\subsection{Implementation}
Benefiting from the auto-configuring capability of nnU-Net, many preprocessing, postprocessing, and architecture hyperparameters can be robustly handled by nnU-Net's default strategies. However, to harness the DA capability of our proposed model, the introduction of a domain classifier necessitates careful selection of additional hyperparameters, along with extra effort to tailor current default segmentation configuration strategies for UDA. 

Firstly, the architecture of the domain classifier must be designed and adjusted to accommodate the complexity of nnU-Net. We introduced the same block structure (i.e., double convolutional layers followed by max pooling) as in the backbone of nnU-Net. Two hyperparameters can be freely adjusted to modify the depth of the domain classifier: i) the number of blocks (set to 4 in our experiments); ii) the number of channels for convolutional layers and the final fully-connected (FC) layer (set to 100).
Secondly, as discussed previously, the scheduler for $\alpha$ in the GRL involves three hyperparameters: $e_{min} = 100$, $e_{max} = 350$, and $\alpha_{max}=3$. Additionally, we observed that the default deep supervision (DS) strategy in nnU-Net tends to prioritize the segmentation task, potentially bypassing the learning of domain-invariant features and leading to an imbalance between segmentation and classification tasks. Therefore, we disabled DS in our proposed model training.
The hyperparameter $\lambda$ is set to 0.01, the batch size is set to 4, the maximum number of epochs is set to 500, and a domain-balanced dataloader is utilized to prevent failure in domain classification. Finally, the learning rate is implemented following the same approach as described in \cite{ganin2016domain}, with an initial value of 0.01.

\begin{table}[t]
\caption{Comparison of methods using 5-fold cross-validation (5CV) (excluding models 1 and 1$^*$).
The models are ranked separately based on whether they have access to adult source data for training, or if deep supervision (DS) was deactivated (marked with $^*$). The best in each category is highlighted in bold.}\label{tab:SDA}
\centering
\begin{tabular}{cccc|ccc|ccc|ccc}
\hline
\multirow{2}{*}{\textbf{Model}} & \multicolumn{3}{c|}{\textbf{mean DSC}}           & \multicolumn{3}{c|}{\textbf{mean HD95}}         & \multicolumn{3}{c|}{\textbf{median DSC}}         & \multicolumn{3}{c}{\textbf{median HD95}}        \\ \cline{2-13} 
                                & ET             & TC             & WT             & ET              & TC            & WT            & ET             & TC             & WT             & ET             & TC            & WT            \\ \hline
1                               &  \textbf{0.583}& 0.522& 0.900&  \textbf{58.77}& 36.11& 4.82&  \textbf{0.742}& 0.549& 0.913&  \textbf{4.12}& 13.49& 3.00\\
2                               & 0.547&  \textbf{0.872}& 0.897& 62.70&  \textbf{5.53}& 4.83& 0.708& 0.915& 0.924& 6.40&  \textbf{3.61}& 3.16\\
3                               & 0.575& 0.854&  \textbf{0.908}& 66.18& 9.38&  \textbf{4.22}& 0.716&  \textbf{0.918}&  \textbf{0.927}& 4.36&  \textbf{3.61}&  \textbf{2.83}\\ \hline
4                               & 0.579& 0.874& 0.898& 66.30& 8.78& 4.55& 0.713&  \textbf{0.914}&  \textbf{0.922}& 5.39&  \textbf{3.32}&  \textbf{3.00}\\
5                               & 0.577& 0.852& 0.902& 52.66& 6.42& 4.66&  \textbf{0.732}& 0.895& 0.917& 5.00& 4.47&  \textbf{3.00}\\
6                               &  \textbf{0.584}&  \textbf{0.879}&  \textbf{0.906}&  \textbf{51.79}&  \textbf{5.23}&  \textbf{4.26}& 0.712& 0.910&  \textbf{0.922}&  \textbf{4.24}& 3.61&  \textbf{3.00}\\
7                               & 0.582& 0.869& 0.901& 55.25& 5.54& 4.31& 0.702& 0.906& 0.920& 4.47& 3.74&  \textbf{3.00}\\
8                               & 0.574& 0.875& 0.903& 62.23& 5.65& 4.54& 0.718& 0.908& 0.921& 5.00& 3.74&  \textbf{3.00}\\ \hline
1$^*$                      & 0.571& 0.526& 0.890& 59.61& 22.25& 7.07&  \textbf{0.739}& 0.555& 0.909&  \textbf{4.47}& 13.49& 3.61\\
2$^*$                       & 0.511& 0.855& 0.886& 50.62& 6.24& 5.37& 0.631&  \textbf{0.909}&  \textbf{0.921}& 7.55&  \textbf{3.74}& 3.16\\
3$^*$                       &  \textbf{0.575}&  \textbf{0.861}&  \textbf{0.909}&  \textbf{45.50}&  \textbf{5.47}&  \textbf{5.16}& 0.721& 0.905& 0.920& 5.10&  \textbf{3.74}&  \textbf{2.83}\\ \hline
\textbf{UDA}             & 0.580& 0.846& 0.892& 50.87& 16.55& 8.90& 0.713& 0.916& 0.923& 4.24& 3.32& 3.00\\ \hline
\end{tabular}
\end{table}

\section{Results and Discussion}
We evaluated the performance of models on $\mathcal{D}_{\text{PED}}^{\text{target}}$ using average and median DSCs along with the 95th percentile Hausdorff distance (HD95).
The results of all models are summarized in Table \ref{tab:SDA}. 
The results revealed that model 1, trained solely on adult data, mainly faces challenges in accurately segmenting the TC region, as evidenced by a decrease in mean DSC from 0.872 to 0.522 compared to model 2. In comparison to model 2, which was trained on a limited dataset of pediatric cases, model 3 benefits from combining both pediatric and adult datasets, resulting in improved median DSC and HD95 metrics due to its exposure to diverse domain characteristics. Since we deactivated the DS in the proposed UDA method, we also trained models 1-3 without DS (models 1$^*$-3$^*$) for fair comparison. As shown, these models perform very similarly to models 1 and 3 with DS with enough training data. Notably, models 3 and 3$^*$ are regarded as practical upper bounds for any DA model. Supervised DA models 4-8 achieve comparable results to the upper bound model trained with both datasets. Lastly, the proposed UDA method obtains segmentation accuracies which are close to the upper bound model 3$^*$ without requiring any annotations from pediatric domain. This underscores the necessity and efficacy of the adaptation.

\begin{table}[t]
\caption{Comparison methods after removing cases with no or small ET region. 
The best values are shown in bold, while the second-best values are underlined.}\label{tab:UDA_remove}
\centering
\begin{tabular}{cccc|ccc|ccc|ccc}
\hline
\multirow{2}{*}{\textbf{Model}} & \multicolumn{3}{c|}{\textbf{mean DSC}}           & \multicolumn{3}{c|}{\textbf{mean HD95}}         & \multicolumn{3}{c|}{\textbf{median DSC}}         & \multicolumn{3}{c}{\textbf{median HD95}}        \\ \cline{2-13} 
                                & ET             & TC             & WT             & ET              & TC            & WT            & ET             & TC             & WT             & ET             & TC            & WT            \\ \hline
1$^*$                      & \underline{0.661}& 0.543& 0.894& 27.50& 14.75& 7.29&  \textbf{0.768}& 0.522& 0.908&  \underline{3.61}& 13.49& 3.74\\
2$^*$                       & 0.616& 0.857& 0.894& 14.05& 6.36& 5.27& 0.729&  0.900&  0.916& 5.10&  4.24& 3.46\\
3$^*$                       &  \textbf{0.685}& \underline{0.863}&  \textbf{0.913}&  \textbf{12.98}&  \textbf{5.74}&  5.46& \underline{0.760}& \underline{0.902}& \underline{0.918}& \underline{3.61}&  \underline{4.12}&  \textbf{3.00}\\ 
\textbf{UDA}             & 0.660& \textbf{0.864}& \underline{0.908}& \underline{13.22}& \underline{5.90}& \textbf{5.15}& 0.757& \textbf{0.913}& \textbf{0.919}& \textbf{3.16}& \textbf{3.74}& \underline{3.16}\\ \hline
\end{tabular}
\end{table}

\begin{figure}[!h]
     \centering
         \centering
         \includegraphics[width=0.48\textwidth]{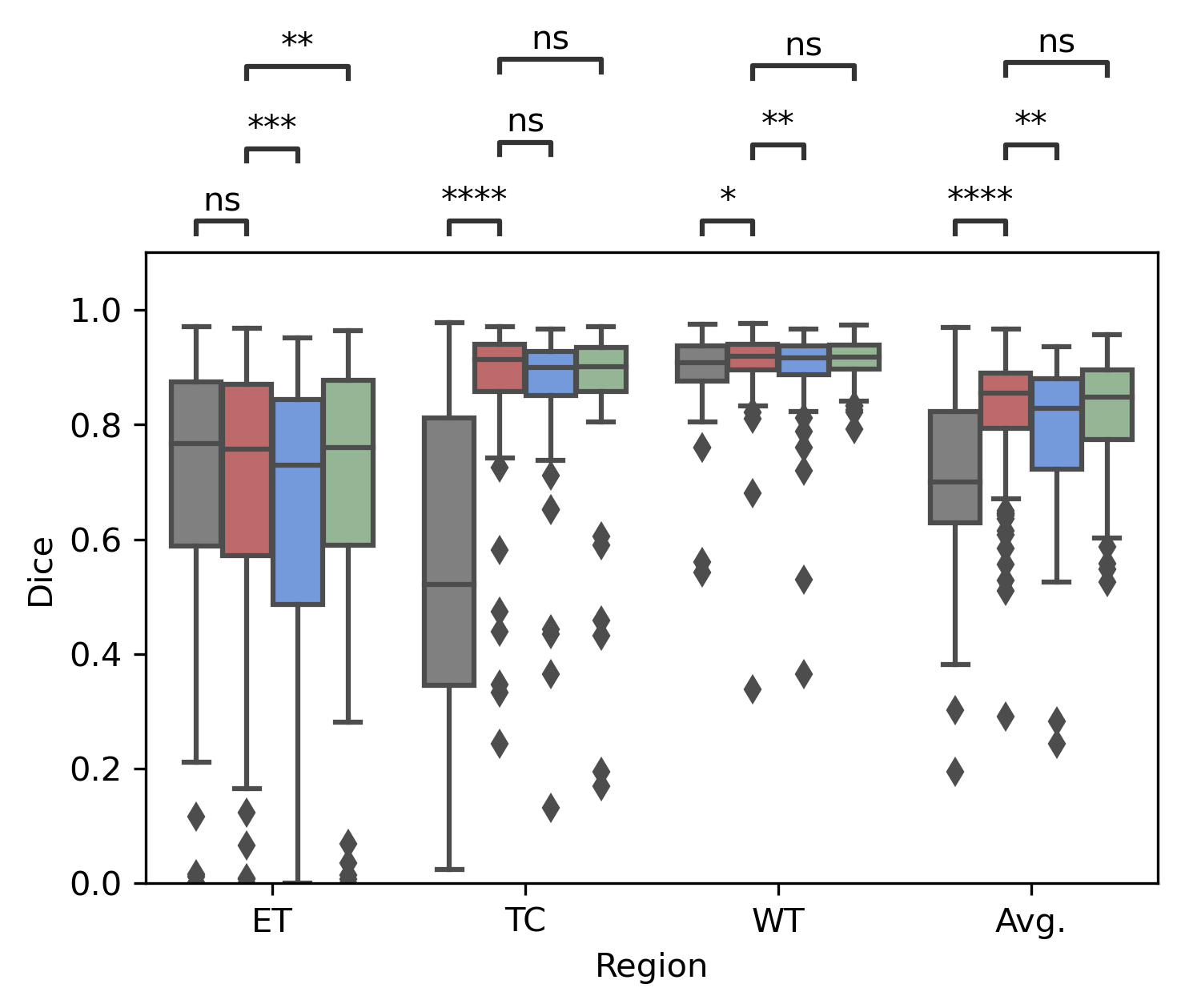}
         \label{fig:results_GRL}
     \hfill
         \centering
         \includegraphics[width=0.48\textwidth]{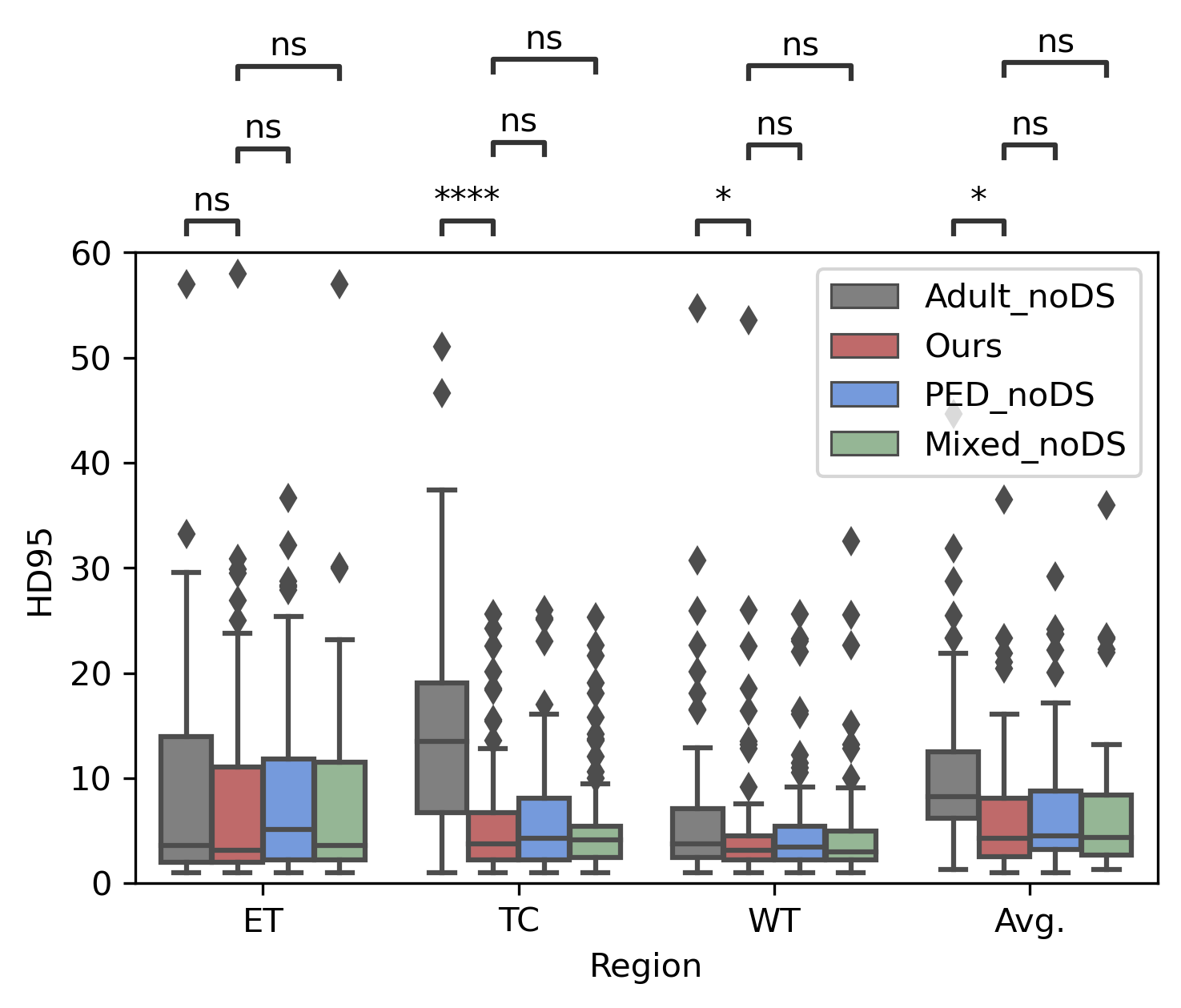}
         \label{fig:visual_illustration}
\caption{Distributions of DSC (left) and HD95 (right) for Table \ref{tab:UDA_remove}. 
Pairwise t-tests were used to assess the statistical significance of differences between methods on $\mathcal{D}_{\text{PED}}^{\text{target}}$. 
} \label{fig:results}
\end{figure}

From the results of Table \ref{tab:SDA}, it is also evident that all tested methods have problems with segmenting ET: relatively low mean DSC and high mean HD95. Notably, there are significant disparities between median and mean metrics, with median values often considerably better than mean values (e.g., 4.12 median vs. 58.77 mean HD95 for model 1). 
As already mentioned, ET regions are usually very small or non-existent in pediatric cases. 
To assess the impact of small ET, we reevaluated models 1$^*$-3$^*$ and our UDA method without considering subjects with ET < 60 voxels (22 subjects were discarded, results shown in Table \ref{tab:UDA_remove}). 
We observe that DSC and especially HD95 are largely improved for all methods in the ET region. This suggests that the considerable variation in ET segmentation performance is largely influenced by small region size. 
Moreover, the proposed UDA method is not statistically different from the practical upper bound model 3$^*$ (see Fig. \ref{fig:results}), except for the most challenging ET region in DSC metric. 



\section{Conclusion}


Our experimental results show that the proposed unsupervised domain adaptation approach performs similarly to supervised approaches for pediatric tumor segmentation. The main advantage of our approach is that it has the potential to be used in cases where expensive manual annotations are not available without a significant drop in performance.

\begin{credits}
\subsubsection{\ackname} This study was partially funded by Barncancerfonden (grant number MT2022-0008). We also thank Berzelius for their computational infrastructure. 

\subsubsection{\discintname}
The authors have no competing interests to declare.
\end{credits}
%
%
%
\bibliographystyle{splncs04}
\bibliography{mybibliography}
%




\end{document}